\documentclass[sn-mathphys,Numbered]{sn-jnl}

\usepackage{graphicx}%
\usepackage{multirow}%
\usepackage{amsmath,amssymb,amsfonts}%
\usepackage{amsthm}%
\usepackage{mathrsfs}%
\usepackage[title]{appendix}%
\usepackage{xcolor}%
\usepackage{textcomp}%
\usepackage{manyfoot}%
\usepackage{booktabs}%
\usepackage{algorithm}%
\usepackage{algorithmicx}%
\usepackage{algpseudocode}%
\usepackage{listings}%

\usepackage{tcolorbox}
\usepackage{wrapfig}
\usepackage[pangram]{blindtext}
\newtcolorbox{mybox}[2]
{colback=#2!5!white,colframe=#2!75!black,fonttitle=\bfseries,title=#1, flushright title, width=\linewidth, height=18cm}

\newcommand{\bu}{\boldsymbol{u}}

\begin{document}

\title[Article Title]{A review of the accuracy of Direct Numerical Simulation tools for the simulation of non-spherical bubble collapses}


\author[1]{\fnm{Mandeep} \sur{Saini}}\email{manndeepsainni@gmail.com}

\author[1]{\fnm{Lucas} \sur{Prouvost}}\email{lucas.prouvost@gmail.com}

\author[1]{\fnm{Stephane} \sur{Popinet}}\email{popinet@basilisk.fr}

\author*[1]{\fnm{Daniel} \sur{Fuster}}\email{daniel.fuster@sorbonne-universite.fr}

\affil*[1]{Sorbonne University, CNRS, Institute Jean Le Rond d'Alembert, F-75005 Paris, France}




\abstract{Numerical methods for the simulation of cavitation processes have been developed for more than 50 years. The rich variety of physical phenomena triggered by the collapse of a bubble has several applications in medicine and environmental science but requires the development of sophisticated numerical methods able to capture the presence of sharp interfaces between fluids and solid/elastic materials, the generation of shock waves and the development of non-spherical modes. One important challenge faced by numerical methods is the important temporal and scale separation inherent to the process of bubble collapse, where many effects become predominant during very short time lapses around the instant of minimum radius when the simulations
are hardly resolved. In this manuscript we provide a detailed discussion of the 
parameters controlling the accuracy of Direct Numerical Simulation 
in general non-spherical cases, where a new theoretical analysis is presented to generalize existing theories on the prediction
of the peak pressures reached inside the bubble during the bubble collapse. We show that the ratio between the gridsize and the minimum radius allow us to scale the numerical errors introduced by the numerical method in the estimation of different relevant quantities for a variety of initial conditions.
}




\maketitle

\section{Introduction}\label{sec1}
Bubbles have attracted the interest of biologists since long ago. One significant area of study emerged during world war, where the absence of pressurized air chambers in aircraft posed a risk of decompression sickness among aviators \cite{harvey1944bubble,harvey1945decompression}. Decompression sickness, also known as 'the bends', occurs when dissolved gases, like nitrogen, form bubbles in the bloodstream during rapid decompression, leading to tissue damage and other symptoms. Over the years scientists have learnt to use bubbles in various applications. These include drug delivery, where bubbles serve as carriers for medications, lithotripsy and histotripsy for breaking down kidney stones and for tissue ablation respectively, High-Intensity Focused Ultrasound (HIFU) treatment for non-invasive treatment, needle-free injection methods, high-contrast ultrasound imaging for medical diagnostics, thrombolysis and many more \cite{alexandrov2004ultrasound,ferrara2007ultrasound,coussios2007role,johnsen2008shock,shekhar2016vitro,roovers2019role,lattwein2020sonobactericide,schoppink2022jet,sieber2023cavitation,shakya2024ultrasound}. In order to gain better control over these biological applications, it is essential to understand the response of these bubbles to the changes in their surroundings. Often in these systems, the bubbles experience a strong pressure difference with respect to their surroundings which leads to a sudden collapse of these bubble \cite{prentice2005membrane,qin2006acoustic,van2006vibrating}. This complex process involves several intricate mechanisms such as high speed liquid jetting, emission of shock waves, light emission \cite{tomita1983collapse,tomita1986mechanisms,brenner2002single,brujan2004role,supponen2017shock,supponen2016scaling}. Additionally, the collapse process is associated with small length and time scales features which makes experimental measurements of several quantities unreliable. To access these quantities and improve understanding of cavitation bubbles, we have to develop reliable numerical codes that are capable of accurately capturing the non-linear response of the bubbles and the consequences of it in its surroundings. \\

We begin by presenting a brief synopsis of historic developments of the numerical methods for studying cavitation. Early development of numerical codes developed to study the cavitation processes are based on the resolution of a Rayleigh--Plesset like equation coupled with a very simple model for the bubble content \cite{hickling1964collapse,lilliston1966calculations,solomon1967nonlinear,prosperetti1988nonlinear,tomita1979effects}. Because an important limitation of these methods is their inability to capture non-spherical effects, Boundary Integral Methods (BIM) were developed in the 70s and 80s as a class of numerical methods capable of simulating non-spherical collapse of bubbles \cite{plesset1971collapse,blake1986transient,blake1987cavitation,wilkerson1990boundary}. Conventionally in BIM methods, the velocity potential is integrated thereby assuming an inviscid and incompressible flows and the pressure at the bubble interface is treated as a stress free boundary condition where the pressure is imposed as 
$$p_{b} = p_{g,0} \left(\frac{V_{g,0}}{V_g}\right)^\gamma,$$
with $V_{g,0}$ the bubble volume at some reference pressure $p_{g,0}$.
This method can be extended to a weakly compressible regimes \cite{wang2011non,wang2013non} being still extensively used for studying the bubble collapse phenomenon \cite{klaseboer2006dynamic,klaseboer2007interaction,li2020modelling,li2021comparison,li20233d}.
Other initial alternatives to BIM include 
few methods proposed to solve the Euler equations for both phases \cite{muller2010numerical,lauer2012numerical}
and a family of viscous solvers that still treated
the interface as a free surface boundary condition \cite{mitchell1973asymmetric,yu1995collapse,popinet2002bubble}.\\

Currently active areas of research involve the development of accurate three dimensional numerical methods with viscous and surface-tension effects while considering evolution of both liquid and gas phase to be able to predict the conditions generated inside the bubble and its surroundings. 
A very popular class of compressible multiphase codes are based on solving the Riemann problem in each cell face \cite{saurel1999multiphase,johnsen2006iws,johnsen2009,schmidmayer2017model,shukla2014nonlinear,tiwari2013diffuse}, which typically resort to the use of Diffuse Interface Methods (DIM) based on the smearing of the interface over several cells. One source of error in these methods is the artificial diffusion of the interface which needs to be limited by special techniques. Johnsen and Colonius \cite{johnsen2009} solved for the compressible Euler's equations for the bubble and liquid. The results showed grid converged results\footnote{when the numerical solution obtained is shown not to depend on the size of the grid elements}  for the major part of a bubble collapse cycle, the bubble displacement velocity and the wall pressure for several standoff distances and for a relatively intense collapse. However, these results
already highlighted the difficulties to 
obtain converged results close to minimum volume.  Tiwari \textit{et al.} \cite{tiwari2013diffuse} also developed a similar method while considering non-equilibrium effects of pressure and velocity across the phases. The results obtained converged to Keller-Miksis model for spherically symmetric collapse. The convergence of the kinetic energy for a non-spherical collapse was also briefly discussed.
Shukla \cite{shukla2014nonlinear} developed an interface sharpening procedure that improved the convergence of spherically symmetric under-water explosion problem.  Phan \textit{et al.} \cite{phan2019numerical} developed a homogeneous mixture model for understanding the bubbles in underwater explosion. For validation the comparison of the bubble temporal radius evolution for the first bubble cycle was compared with the experimental observations. Schmidmayer \textit{et al.} \cite{schmidmayer2017model,schmidmayer2020ecogen,schmidmayer2020assessment} discuss the effect of different models and numerical schemes showing that non-spherical effects can develop during the collapse of bubble in bulk, close to the instant of minimum volume. Interestingly, these effects are shown to be sensitive to the numerical model and advection scheme. For validation they also show radius evolution for different grids and present RMS errors as compared to the Keller-Miksis model.  Panchal \textit{et al.} \cite{panchal2023seven} described a 7 equation model with diffused interface for compressible multiphase flows. 
Paula \textit{et al.} \cite{paula2023robust} described a higher order method for interface capturing and discrete equation method for solving multiphase flow equations. They discussed the grid convergence of a spherically symmetric detonation problem.\\

Another class of methods which have recently gained popularity are based on the extensions of the classical projection methods\footnote{A method to obtain a numerical solution of incompressible Navier Stokes equations which imposes the divergence free condition at the discrete level} of incompressible flows to compressible regimes. A common theme of these methods is the solution of a Poisson-Helmholtz equation for evolution of pressure. Miller \textit{et al.} \cite{miller2013pressure} developed a pressure based method for numerical simulations of bubbles showing the predicted radius evolution for meter sized bubbles for different grids and comparing their results with experiments. Koch \textrm{et al.} \cite{koch2016numerical} developed a pressure based solver using OpenFoam where they showed very good grid convergence for several quantities like collapse time, rebound radius etc. They have used enormous mesh compression near the minimum volume in order to correctly resolve the interface near the minimum volume. Denner \textit{et al.} \cite{denner2020modeling} have also proposed a slightly different pressure based solver where a large system of linear equations is solved simultaneously for pressure and velocity and where the interface between the two fluids is represented with algebraic volume of fluid method. They showed that the bubble evolution converges to Gilmore model for pressure ratio of 25 between the ambient and bubble pressures. They also present results for two different time steps and revealed that the results might deviate from Gilmore model if the time step is not small enough. This method has been able to reproduce well the experimentally measured wall pressure during the collapse a nearby bubble \cite{gonzalez2021acoustic}.  Fuster and Popinet \cite{fuster2018} developed a consistent and conservative all-Mach method for bubble dynamic problems in the Basilisk code where the interface is represented with a geometric Volume Of Fluid method (VOF). They discussed several spherical and non-spherical bubble collapse problems and show that the results converge to the Keller--Miksis solution for the spherical case. Recently, this method is also extended to include the heat transfer effects across the interface \cite{saade2023multigrid}.\\


In this article, we use
the all-Mach method of Fuster and Popinet \cite{fuster2018}
to discuss the influence of numerical errors
on the prediction of relevant quantities associated to the bubble collapse. For that, the Rayleigh collapse problem will be simulated under different initial configurations with variable intensity. The influence of various effects (viscosity, surface tension, non-spherical effects) on the quantities of interest will be discussed from a theoretical point of view, clarifying the length scales that one needs to resolve, in order to ensure accurate results of the Direct Numerical Simulation of the process of bubble collapse.

\section{Classical models and numerical methods for DNS of bubble dynamics}

In this manuscript 
we discuss methods able to solve for the compressible
Navier--Stokes equations in both gas and liquid phase, which
require the resolution of an advection equation to track the position of the
interface (for example the fraction of a reference fluid $f$)
$$ \frac{\partial f}{\partial t } + \bu \cdot \nabla  f = 0,$$
and a set of conservative equations
  \begin{eqnarray}
   && \frac{\partial Y_i}{\partial t } + \nabla \cdot \boldsymbol{F}_i  = 0
      \end{eqnarray}
      where $Y_i=(\rho_i f_i,\rho_i f_i \bu, f_i \rho_i e_{T,i} )^T$ denotes the conservative variables (density, momentum and total energy)
      associated to the gas and liquid phase and $\boldsymbol{F}_i$ is its corresponding flux
$$
~~~\boldsymbol{F}_i = 
\begin{pmatrix}
        f_i \rho_i \boldsymbol{u}_i\\
        f_i \rho_i \boldsymbol{u}_i  \boldsymbol{u}_i - \boldsymbol{\tau_i} \\
        f_i \rho_i e_{T,i} \boldsymbol{u}_i - f_i (\boldsymbol{\tau}_i \cdot \boldsymbol{u}_i - \boldsymbol{q}_i)
\end{pmatrix},
$$
with $\boldsymbol{\tau}$ the stress tensor and $\boldsymbol{q}_i=-\kappa_i \nabla T_i$ the diffusive heat flux. The system above is closed by adding an Equation Of State (EOS).
Different methods
have been proposed for the solution of the system of equations above \cite{fuster2019review}.
Some methods reformulate the system of equations above 
using non-conservative primitive variables. For example, it is possible to 
obtain an evolution equation for pressure from the basic conservation equations as \cite{fuster2018}
\begin{eqnarray}
\label{eq:peq}
    \frac{1}{\rho c_{\rm eff}^2}{D p \over Dt} - {\beta_T \Phi_v \over \rho c_p} = - \nabla \cdot \bu,
\end{eqnarray}
where $\Phi_v$ denotes the viscous dissipation and $\frac{1}{\rho c^2}\bigg|_{\rm{eff}} = \frac{\gamma}{\rho c^2} - \frac{\beta_T^2 T}{\rho c_p}$ is a thermodynamic property
which is equal to $\frac{1}{\rho c^2}$ for perfect gases ($\beta_T=1/T$; $\gamma=C_p/C_v$) and
also weakly compressible liquids ($\beta_T=0$, $\gamma=1$).
In addition, we can use the internal energy equation to write
an explicit equation for the fluid temperature evolution as
\begin{equation} \rho_i c_{p,i}\frac{DT_i}{Dt} = \beta_i T_i \frac{Dp_i}{Dt} + \Phi_v - \nabla\cdot\boldsymbol{q}_i.
\end{equation}

The choice of the primitive variables is arbitrary but has an 
important impact on the conservation properties of the method and therefore
how the numerical errors are going to impact the numerical results. 
.\\

Various challenges are faced when solving
the multiphase compressible equations in the context of bubble collapse. First, for very violent collapses the solver must be able to capture the appearance of shock waves in the liquid that are eventually responsible of a significant energy exchange of the bubble with its surroundings. At the same time, the amplitude of the wave quickly decays and the propagation of waves responsible for noise emission become linear. Thus, numerical methods need to reduce the damping introduced on the propagation of acoustic waves (low Mach number) while being able to correctly capture the emission of shock waves during the bubble collapse.\\

Another problem associated to solvers based on the coupled solution of the equations both inside and outside the bubble is how to deal with the discontinuity of pressure across the interface imposed by the Laplace equation
\begin{equation}
\label{eq:laplace}
p_1 = p_2 - \sigma \kappa - \mu_2 2 \boldsymbol{n} \boldsymbol{D}_2 \boldsymbol{n} + \mu_1 2 \boldsymbol{n} \boldsymbol{D}_1 \boldsymbol{n}, ~~~~~ \boldsymbol{x}=\boldsymbol{x}_I.
\end{equation}
The ability of the method to handle discontinuities on the pressure field
poses various technical difficulties. Spurious parasitic currents due to curvature error computation
are typically discussed in static configurations \cite{popinet2009accurate}, but these test do not provide
information about capability of the method to capture the normal stress viscous jump. The errors introduced in the discretization of interfacial cells usually have 
an important impact on the development of physically meaningful interfacial instabilities
which can be eventually damped by physical but also numerical diffusion effects. 
This complicates the discussion about the appearance of non-spherical modes when 
comparing two numerical methods in unstable configurations, as an optimal numerical method 
should minimize the generation of spurious numerical effects while keeping a relatively small viscous dissipation. The problems associated with the discretization of the interface
typically appear for very refined grids, when the viscous dissipation is not sufficient to damp
the artificial discretization errors introduced. Thus, a naive analysis of multidimensional
solvers based on the capability of the solver to reproduce an spherically symmetric collapse is misleading, as very dissipative methods can
provide the wrong impression that they are accurate just because they damp non-spherical modes.\\

A less common issue discussed with the methods mentioned above is the problem
of entropy conservation. In many models, the bubble response is assumed to be adiabatic and therefore the entropy generation should be zero. From the theorem of Zhong and Marden \cite{zhong1988lie}, a numerical scheme
cannot simultaneously preserve momentum, energy, and symplecticity but
only two out of the three \cite{llor2019geometry,vazquez2017novel}.
For numerical methods solving for the discretized pressure equation
\ref{eq:peq}
    this effect can be easily seen for an ideal gas
by computing the term ${1 \over \rho c^2} {D p \over Dt}$ as  
    $${1 \over \rho c^2} {D p \over Dt} = {1 \over \gamma} {D \over Dt} \left( \ln \left( \gamma p \right) \right) = 
    {1 \over \gamma} { \ln \left( \gamma p^{n+1} \right) -  \ln \left( \gamma p^* \right) \over \Delta t}  =
    {1 \over \Delta t \gamma}  \ln { p^{n+1} \over p^*}$$
    and the right hand size as
    $$\nabla \cdot \boldsymbol{u} = -\frac{1}{\rho}  {D \rho \over Dt} = -{D  \over Dt}(\ln(\rho)) = \frac{-1}{\Delta t}\ln\left(\frac{\rho^{n+1}}{\rho^*}\right)$$
    where we have used that $\rho c^2 = \gamma p $
    and $p^*$ and $\rho^*$ denotes the gas pressure and density at $t_n$ of the fluid particle following a Lagrangian trajectory. 
    While in the continuum limit we naturally obtain that the quantity
    $$\frac{p^{n+1}}{(\rho^{n+1})^\gamma}=\frac{p^{*}}{(\rho^{*})^\gamma}$$
    remains constant for a Lagrangian fluid particle, this condition is not respected at the discrete level.
    Thus, the classical discretization of 
    the left hand side of eq. \ref{eq:peq}
    is obtained using Taylor series to develop the logarithmic term. Neglecting the viscous contribution we obtain
    $$\ln { p^{n+1} \over p^*} 
    = \sum_{k=1}^\infty \frac{(-1)^{k+1}}{k} \left({ p^{n+1} -p^* \over p^*} \right)^k  =  { p^{n+1} -p^* \over p^*} + \mathcal{O}(\Delta t^2)$$
    while the right hand size is discretized as
    $$\nabla \cdot \boldsymbol{u} = \nabla \cdot \boldsymbol{u}^*  - \nabla \cdot \left( \frac{\Delta t}{\rho} \nabla p \right)$$
    with
    $$ p^* =  p^n - (\boldsymbol{u} \cdot \nabla p)^n \Delta t + \mathcal{O}(\Delta t^2),$$
    $$ \boldsymbol{u}^* =  \boldsymbol{u}^n - \nabla \cdot ( \boldsymbol{u}\boldsymbol{u}))^n \Delta t + \mathcal{O}(\Delta t^2).$$
We can realize that standard discretization schemes drop $\Delta t^2$ terms introducing errors on 
the discrete entropy conservation due to temporal discretization errors. Entropy production issues are associated with the solvers ensuring mass, momentum and energy conservation and in general to any unconditionally stable solver, where entropy production is needed to ensure stability. This fact implies that even in discretely conserving methods, the numerical errors are translated into an effective dissipation that break the isoentropic condition usually adopted to simulate the collapse of bubbles. The necessity of quantifying the influence of numerical errors on the artificial production of entropy by the numerical method
is of crucial importance to accurately predict the peak pressures during the collapse and the strength of emitted waves, but it is an aspect that remains practically unexplored in the current literature.\\

\section{Measurable quantities, reference solutions and convergence analyses for code validation}\label{sec3}

The evolution of Computational Fluid Dynamics (CFD) codes requires the validation of the correct treatment of the various terms
included. A classical benchmarking test case is the Rayleigh collapse
problem\footnote{This problem, posed by Rayleigh in 1917, consists in determining the temporal evolution of the bubble volume of a bubble collapsing in a liquid bulk by the difference between the bubble and the ambient pressure\cite{rayleigh1917viii}}, in which a bubble at some initial pressure $p_{g,0}$ is initialized in a liquid at rest and at a higher ambient pressure $p_\infty$. 
The generalization of this test to systems where the bubble is initially close or in contact to a solid wall \cite{saini2022dynamics} or bubble clusters \cite{rossinelli201311} is interesting because it captures all the physical effects typically encountered
in cavitation phenomena in real applications including the emission of shock waves, the appearance of liquid jets, etc. 
In the following, we will discuss the reference solutions used to 
compare the numerical results obtained in order to validate a numerical code. \\

In the spherically symmetric case, viscous and surface tension effects are trivially added in the Rayleigh-Plesset equation to provide an accurate solution that can be used as reference for code validation. In turn, compressibility effects
can be only partly accounted for using different modifications of the original Rayleigh--Plesset model 
which assume that the acoustic wave emitted is linear, 
restricting the applicability of such models to small Mach number regimes \cite{keller1980bubble,gilmore1952growth,tomita1979effects}. The introduction of compressibility effects
in the full problem
introduces an extra degree of freedom when specifying the initial conditions for the liquid pressure and density fields. 
Several initial conditions including that of a sudden jump of pressure at the interface are in principle admissible. 
However, because the models proposed in \cite{keller1980bubble,gilmore1952growth,tomita1979effects} are just first order corrections of the incompressible solution it is  convenient to initialize the pressure field 
from the solution limit of an incompressible liquid if one wants to compare the solutions of the solver with these simplified models.
The spatial structure of the initial pressure field has analytical solution in the case of an spherical bubble, while in the case of bubbles that are not initially spherical or where the domain is not infinite it is required to obtain it numerically. Assuming that the liquid is an incompressible substance, 
we can solve a Laplace equation for the pressure in the liquid domain with 
Dirichlet boundary conditions at the interface determined
by the Laplace equation \cite{saini2022dynamics}.
This will be the procedure to obtain the initial condition in the simulations shown along this manuscript. \\

The Rayleigh collapse problem is a challenging test case when looking at instants near the minimum bubble volume. Especially for strong collapse intensities, 
the bubble interface near the minimum volume is often hard to resolve as the bubble can shrink by at least one order of magnitude in size. In addition, the axi-symmetric and fully three-dimensional problem is unstable and 
external non-spherical perturbations are amplified due to
the presence of physically meaningful Rayleigh--Taylor instabilities leading to length scales which are much smaller than
the minimum bubble radius.\\

The large scale dynamics of the bubble can be described using the inherent characteristic velocity
of the problem $U_{c,1} = \sqrt{p_\infty/\rho_l}$ 
and the initial bubble radius $R_0$.
For example, the dimensionless collapse time $\frac{t_c U_{c,1}}{R_0}$ for an spherical bubble
found by Rayleigh can be written as
\begin{equation}
    \frac{t_c U_{c,1}}{R_0} = 0.915  \sqrt{1 -\frac{p_{g,0}}{p_\infty}},
    \label{eq:tcollapse}
\end{equation}
which becomes an $\mathcal{O}(1)$ quantity for sufficiently large values of $p_\infty/p_{g,0}$.
Based on these scalings, the relevance of viscous and surface tension effects are expected to scale with the following definitions of the Reynolds and Weber number
\begin{eqnarray}
  \rm{We}_0 = \frac{p_\infty R_0}{\sigma}, ~~~~ \rm{Re}_0 = \frac{\sqrt{p_\infty \rho_l} R_0}{\mu_l}. 
\end{eqnarray}
These dimensionless quantities naturally appear in the
solution of a linear (weak) oscillation of a bubble in a free liquid \cite{fuster2015mass,bergamasco2017oscillation}.
The importance of liquid compressibility effects deserves particular attention in strongly nonlinear regimes. For sufficiently intense collapses $U_{c,1}$ is not representative of the flow velocity at the instant close to the collapse, when liquid compressibility effects become important. Instead, we need to use the expression for the peak pressure and minimum volume reached by the collapse of a
gas bubble in an inviscid liquid \cite{Brennen}
\begin{equation}
    \frac{p_{max}}{p_{g,0}} \approx \left({(\gamma - 1) \frac{p_{\infty}}{p_{g,0}}}\right)^{\frac{\gamma}{\gamma - 1}}, ~~~~
    \frac{R_{min}}{R_0} = \left(\frac{p_{g,0}}{p_{max}}\right)^{\frac{1}{3\gamma}},
    \label{eq:pinfsph}
\end{equation}
to introduce a characteristic velocity $u_{max} = \sqrt{p_{max}/\rho_l}$ 
which defines a Mach, Weber and Reynolds number as
\begin{eqnarray}
    \rm{Ma}_{\rm{max}} = \frac{u_{max}}{c_l} ,~~~~ \rm{We}_{\rm{max}} = \frac{\rho_l u_{max}^2 R_{min}}{\sigma}, ~~~~ \rm{Re}_{\rm{max}} = \frac{\rho_l u_{max} R_{min}}{\mu_l}. 
    \label{eq:Mamax}
\end{eqnarray}
It is interesting to note that the maximum values of the Mach, Weber and Reynolds number increases with $\frac{p_{\infty}}{p_{g,0}}$, which implies that liquid compressibility effects become increasingly important as the intensity of the collapse increases while liquid viscosity and surface tension effects are negligible for very strong collapses despite the fact that the bubble becomes small.\\

\begin{table}[]
    \centering
      \caption{Estimation of the maximum values of the ratio between the minimum radius and the minimum gridsize $R_{min}/\Delta x_{min}$ used in different convergence studies reported in the literature for the Rayleigh collapse problem and for underwater explosions. No convergence study on the peak pressures genereated has been found for the shock/bubble interaction problem \cite{hu2006conservative,johnsen2009,lauer2012numerical,paula2023robust} .}
    \begin{tabular}{cccc}
     Reference  (year) &  Rayleigh collapse &  Underwater explosion  \\
    \hline\\
    \cite{luo2004computation} (2003) & & 847   \\
    \cite{hu2006conservative}  (2006) & & 12.8      \\
    \cite{johnsen2009} (2009) & 2 &  \\
   \cite{miller2013pressure}  (2013) &  & 33    \\
    \cite{shukla2014nonlinear} (2014) & & 80   \\
    \cite{koch2016numerical} (2016) & 6   \\
    \cite{fuster2018} (2018) & 40  \\
     \cite{wermelinger2018petascale} (2018) & 73  \\
    \cite{lechner2020jet} (2020) & 20  &  \\
    \cite{denner2020modeling} (2020) & 54 &  \\ 
    \cite{schmidmayer2020assessment} (2020) & 15 &  \\
    \cite{li2021comparison} (2021) & & 1000  \\
    \cite{nguyen2022modeling} (2022) & 3 &  \\
    \cite{park2022numerical} (2022) & 10 &  \\
    \cite{paula2023robust} (2023) & & 65  \\
    \cite{duy2023numerical} (2023) & & 20  \\
    \cite{yang2023numerical} (2023) & 2  
    \end{tabular}
  
    \label{ConvergenceStudies}
\end{table}

Schmidmayer et al \cite{schmidmayer2020assessment} clearly shows that the errors introduced 
in the simulation are especially important at the instant of minimum radius, where the peak pressures are reached, the errors being also sensitive to the numerical method chosen.
An estimation of how the minimum radius depends on the pressure ratio
can be obtained as $$\frac{R_{min}}{R_0} \approx 2 \left(
 \frac{p_{g,0}}{p_\infty} \right)^{\frac{1}{3(\gamma-1)}}.$$ 
 Thus, simulations for large values of the ratio 
 $\frac{p_\infty}{p_{g,0}}$ are very challenging
 if one requires to solve the physical phenomena taking place at the collapse.
 Table \ref{ConvergenceStudies} shows a list of numerical works
 where the influence of the grid size on the accuracy of the numerical predictions is discussed. In addition to the Rayleigh collapse problem,
 we also include two related tests: the underwater explosion test and
 the problem of shock/bubble interaction.
 The problem of underwater explosion is indeed analogous to the Rayleigh collapse problem except that the simulation is initialized at the instant
of minimum radius (maximum pressure) and therefore the initial spatial discretization controls the maximum errors introduced in 
the simulation.
In the problem of shock/bubble interaction, the ratio between the pre-shock
and post-shock pressure can be taken as a measurement of the bubble collapse strength.
We can clearly see that in the case of underwater explosions all the authors reporting convergence studies use at least 10 points per radius in order to 
limit the sensitivity of the numerical predictions to the grid size.
In the case of the Rayleigh collapse problem, most of the authors reporting convergence studies also use at least 10 points per radius at the instant of minimum volume, although it is possible to find studies with lower resolutions where the authors limit themselves to showing the temporal evolution of the bubble volume and not the peak pressures reached at the collapse.  Finally,
although it is possible to find many numerical studies of the shock/bubble interaction problem \cite{hu2006conservative,johnsen2009,lauer2012numerical,paula2023robust} (just to mention few),
to the best of our knowledge no grid convergence studies have been reported
for this problem up to date. One possible reason for this is that most authors use this test to validate the codes by reproducing experimental conditions where the pressure ratios are extremely large (typically of the order of $10^3$). Thus, although simulations are well resolved during the first instants when the shockwave deforms the bubble and the methods capture well the patterns experimentally observed, 
the capability of numerical simulations to resolve the  peak pressures reached during the collapse is beyond currently available computational resources.\\

\begin{figure}[h!]
    \centering
    \includegraphics[scale = 0.85]{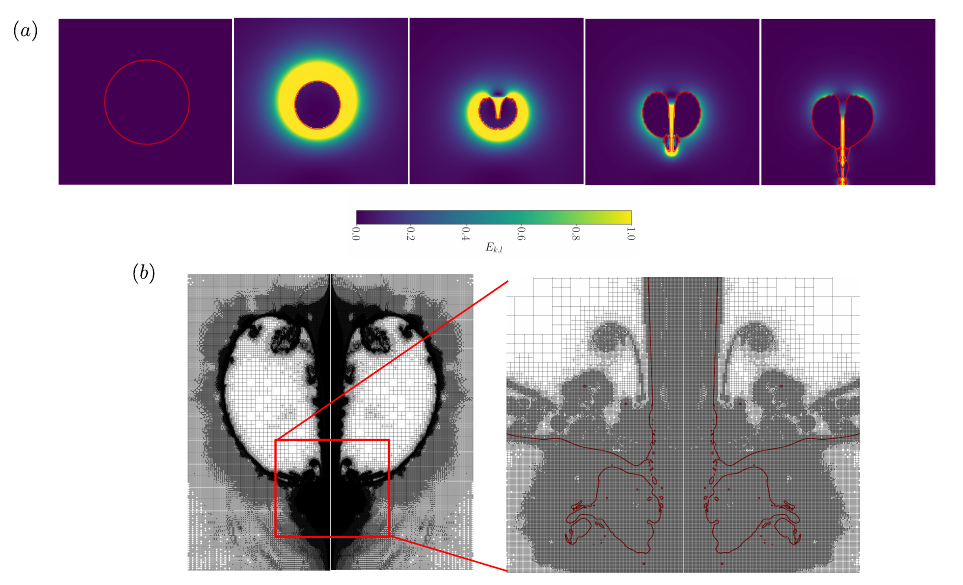}
    \caption{$(a)$ Interface isocontour and kinetic energy distribution for collapse of 0.5 mm bubble under the influence of $p_\infty/p_{g,0}=10$ at distance $d=2 R_0$ from the no-slip wall for $\textrm{Re}_0 = 5000$ and $\textrm{We}_0 = 695$, $\rm{Ma}_{\rm{max}} = 0.085$. $(b)$ Zoomed in view of the grid size distribution at the instant of liquid jet development near the fourth frame from left in the top panel. }
    \label{fig:bubbleclosewall}
\end{figure}

 \begin{figure}[hbt!]
  \centering
    \includegraphics[scale = 0.75]{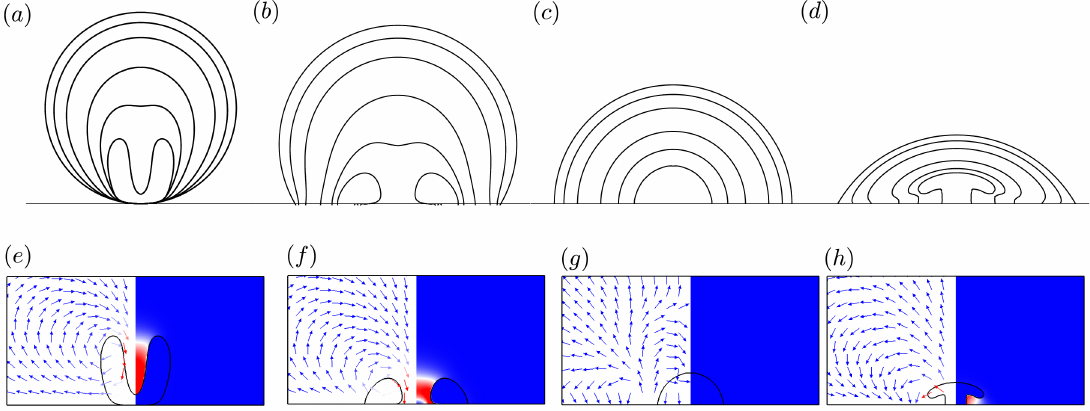} 
    \caption{The DNS results for collapse of spherical cap bubbles attached to a slip wall with different contact angle $\alpha$ and for $p_\infty/p_{g,0} = 8$, $\rm{Re}_0 = \infty$, $\rm{We}_0 = \infty$. The evolution of bubble shapes contours for $(a)$ $\alpha = 0$ $(b)$ $\alpha = \pi/3$ $(c)$ $\alpha = \pi/2$ $(d)$ $\alpha = 2/3\pi$. 
    }
\label{fig:evol}
\end{figure}

\begin{figure}[h!]
    \centering
   \begin{tabular}{cc}
   $(a)~~~~~~~~~~~~~~~~~~~~~~~~~~~~~~~~~~~~~~~~~~~~~~~~~~~(b)~~~~~~~~~~~~~~~~~~~~~~~~~~~~~~~~~~~~$\\
    \includegraphics[scale = 0.65]{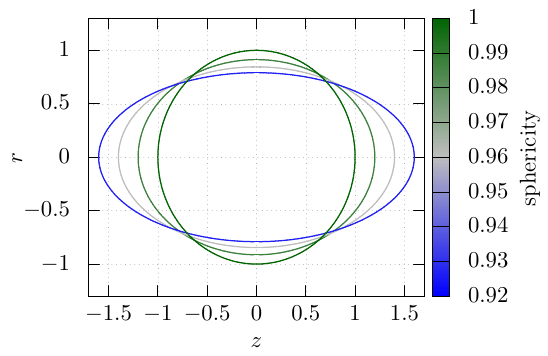}
    \includegraphics[scale = 0.65]{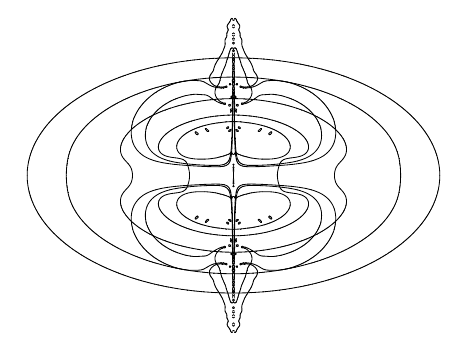}
    \end{tabular}
    \caption{$(a)$ Initial shape of the prolate bubbles used to investigate the influence of non-spherical bubble collapses.
    $(b)$ Interface isocontours of the collapse and rebound of a prolate bubble for $p_\infty/p_{g,0} = 6$, $\rm{Re}_0=10^4$ 
    and $\rm{We}_0=\infty$. }
    \label{fig:prolatebubbles}
\end{figure}

 In the following, we discuss the influence of grid resolution on different quantities associated to the collapse of a bubble
 using the solver proposed in \cite{fuster2018} as a representative example to discuss the importance of numerical errors. 
 In addition to simulations where we enforce spherical symmetry, the solution of
 three different axisymmetric problems will be considered: (i) the Rayleigh collapse of a spherical bubble that can be near a solid no-slip wall (figure \ref{fig:bubbleclosewall}), (ii)  the Rayleigh collapse of a spherical cap bubble in contact with a slip wall with some initial contact angle $\alpha$ (figure \ref{fig:evol}) and (iii)  the Rayleigh collapse of a prolate bubble defined by
 the initial sphericity $\Psi$ of the spheroid defined as
\begin{align}
    \Psi = \frac{\pi^{1/3}\left(6\, V_{b,0}\right)^{2/3}}{S_{I,0}} \nonumber
\end{align}
with $S_{I,0}$ is the interfacial area at t=0 (figure \ref{fig:prolatebubbles}).
In each case, the relative error $\epsilon(u)$ for a given quantity $u$ will be computed as 
$$
    \epsilon(u) = \frac{u-u_{ref}}{u_{ref}}
$$
 where $u$ is the solution obtained with a given resolution and $u_{ref}$ will be the reference solution obtained with the finest possible grid.\\
 
\subsection{Rayleigh collapse time}

\begin{figure}[h!]
    \centering
    \begin{tabular}{cc}
    \includegraphics[scale = 0.9]{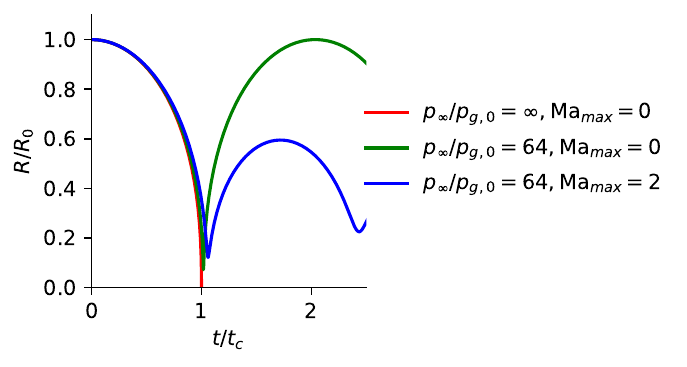}
    \end{tabular}
    \caption{The evolution of bubble radius given by the Rayleigh model $p_\infty/p_{g,0} = \infty, \textrm{Ma}_{\rm{max}} = 0$, the Rayleigh-Plesset model $p_\infty/p_{g,0} = 64, \textrm{Ma}_{\rm{max}} = 0$, and the Keller-Miksis model $p_\infty/p_{g,0} = 64, \textrm{Ma}_{\rm{max}} = 2$.
    In all cases $\rm{Re}_0=\infty$ and $\rm{We}_0=\infty$.}
    \label{fig:tcollapse}
\end{figure}

\begin{figure}[h!]
    \centering
    \includegraphics[scale=0.9]{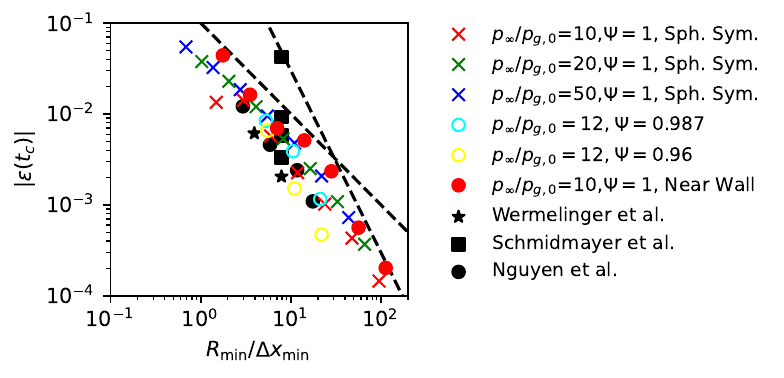}
    \caption{Convergence
    test on the collapse time as a function of the dimensionless grid size defined with the minimum bubble radius. 
    The results are reported for different problems: 
    a spherically symmetric bubble for three different collapse intensities, two different prolate bubble for $p_\infty/p_{g,0} = 12$
    and the axi-symmetric collapse of a 0.5mm air bubble in water near a no-slip wall for $p_\infty/p_{g,0} = 10$, $\textrm{Re}_0 = 5000$ and $\textrm{We}_0 = 695$, $\rm{Ma}_{\rm{max}} = 0.085$. An estimation of convergence errors from references \cite{johnsen2009,wermelinger2018petascale,schmidmayer2020assessment,nguyen2022modeling} are also shown with black symbols.}
    \label{fig:tcollapseconvergence}
\end{figure}


 Generally, equation \ref{eq:tcollapse} is a very good estimate of the collapse time as the corrections due to various factors including viscous effects and  liquid compressibility are fairly small. Figure \ref{fig:tcollapse}  shows that in spherically symmetric models the collapse time does not vary significantly when using different models irrespective whether the liquid is considered incompressible or compressible.  
 As shown in figure \ref{fig:tcollapseconvergence} the errors in this quantity introduced by the Direct Numerical Simulations (DNS) of the Navier--Stokes equations reach small relative errors of the order $10^{-3}$ when we have around $10$ grid points across the minimum bubble radius irrespective whether spherical symmetry is imposed or not, the accuracy of the results being acceptable even for relatively poor resolutions at the instant of minimum radius. The estimate of the errors on the estimation of the collapse time of other works in the literature showing grid convergence results are consistent with the errors reported. \\
  
\begin{figure}[h!]
\centerline{\includegraphics{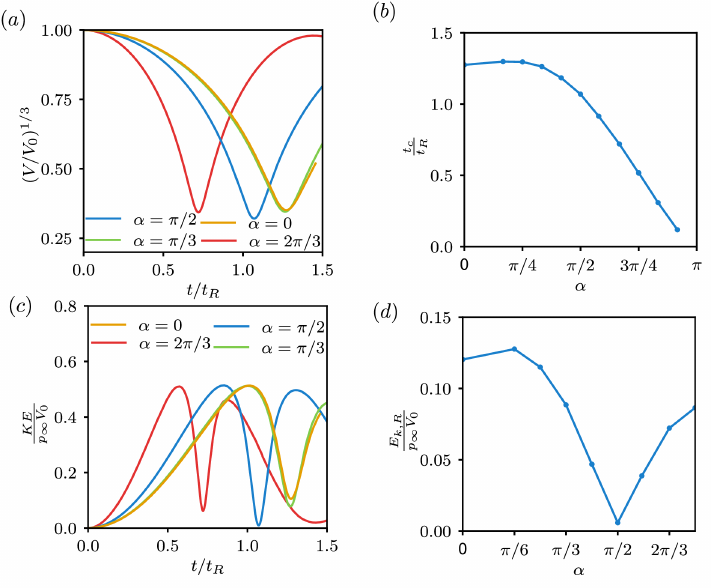}} 
  \caption{Axi-symmetric DNS results for bubbles attached to a slip wall $p_\infty/p_{g,0} = 8$,$\textrm{Re}_0 = \infty$, $\textrm{We}_0 = \infty$  and varying contact angle. $(a)$ Evolution of dimensionless equivalent radius of the bubble. $(b)$ The dimensionless collapse time is plotted as function of $\alpha$. }
\label{fig:vkevst}
\end{figure}

 Many studies have shown converged results for the collapse time for a spherical bubble and also for non-spherical bubbles \cite{koch2016numerical,denner2020modeling,park2022numerical}. In this later case,
 the presence of nearby walls can induce a shielding effect which  leads to a prolongation of the collapse time \cite{reuter2022rayleigh}.
In figure \ref{fig:vkevst}$a$, we show the evolution of non-dimensional equivalent bubble radius for spherical cap bubbles initially in contact with 
a slip wall. The initial curvature is $R_0=1$ and 
$p_\infty/p_{g,0}=8$ and the initial contact angle $\alpha$ varies between $0$ and $2\pi/3$. The minimum of the bubble volume
is only weakly linked with the bubble shape whereas the non-dimensional time of bubble collapse depends significantly on $\alpha$. The bubble collapse time (see figure \ref{fig:vkevst}$b$) for hemispherical bubble compares well with the Rayleigh collapse time: for bubbles with $\alpha < \pi/2$, it is relatively constant and around $1.25 t_R$, and for $\alpha > \pi/2$ it decreases sharply. The prolongation in the bubble collapse time compared to the Rayleigh collapse time is discussed in detail by Reuter et. al. \cite{reuter2022rayleigh} for laser generated bubbles. Interestingly, they also report similar factor of around $1.2$ for $0.5 < d/R_{max} < 1$ (where d is standoff distance for laser focus and $R_{max}$ is the maximum bubble radius), even though the setup and shapes of bubbles are very different. The influence of the contact angle on the collapse time can be 
estimated from the impulse theory, which provides the initial acceleration field. The averaged acceleration along the interface eventually provides an estimation of the changes of the collapse time with respect to the spherical case \cite{sainithesis}.\\

It is remarkable to see in figure \ref{fig:tcollapseconvergence} that
 the errors introduced in different configurations
 for different collapse intensities and initial sphericity scale well with the minimum
 radius and not the initial bubble radius irrespective of the initial shape of the bubble.\\



\subsection{Peak pressures inside the bubble}
The collapse time $t_c$ depends very little on the details of the model and it does not provide important information about the correctness of the method to capture the peak gas pressures.
In the limit of an adiabatic compression, the maximum bubble pressure is reached at the instant of minimum volume, determining the peak gas pressures reached during the bubble collapse. 
Unfortunately, this quantity is not measurable from experiments so the validation of codes
relies on the capability to obtain theoretical estimates. As explained before, an estimation 
of the peak pressures reached by the collapse of a bubble in an incompressible and inviscid liquid 
have been reported by Brennen \cite{Brennen}.
In order to evaluate the sensitivity of this quantity to parameters such as the liquid compressibility, viscosity, or bubble deformation, we can 
resort to energy conservation principles to relate the net rate of change of total system energy ($E$), which is equal to summation of the rate of work done ($\dot{W}_{ext}$), heat transfer ($\dot{Q}_{in}$), and energy dissipation ($\dot{E}_{dis}$) due to irreversible processes
\begin{equation}
\frac{dE}{dt} = \Dot{W}_{ext} + \Dot{Q}_{ext} - \Dot{E}_{dis},
\end{equation}
\begin{figure}[h!]
    \centering
    \includegraphics[]{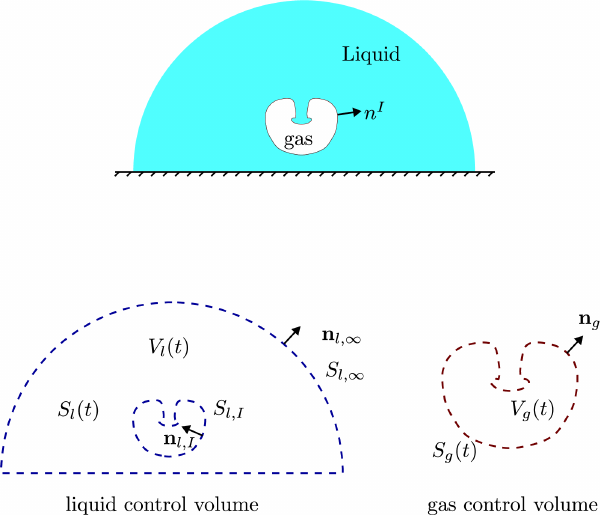}
    \caption{Sketch of the control volumes chosen for the energy analysis of the Rayleigh collapse problem.}
    \label{fig:energy}
\end{figure}
where $E_{dis}$ is always a positive quantity. 
Among the various energies in the system we can identify 
the kinetic energy $E_{k} = \int_{V} \frac{1}{2} \rho \mathbf{u}^2 dV,$
the internal energy associated to the liquid compressibility \cite{Landaubook}
$E_{e,l} = \int_{V_l} \frac{1}{2} \frac{(p_l - p_0)^2}{\rho_{l,0} c_l^2} d{V},$ the internal energy of the gas phase
$     \int_{V_g} \rho_g e_g dV 
     = \frac{p_{b} V_b}{\gamma-1}$,
      the surface energy 
${E_s=\sigma (S_g-S_{g,0})}$ and, if present, the energy associated to the contact line.
Following the developments indicated in Appendix \ref{secA1}, energy conservation principles can be used to obtain  an equation  
that relates the liquid elastic energy, the surface energy and viscous dissipation through the definition of a potential energy that depends on the bubble volume
\begin{equation}
    \frac{E_{k,l} + E_s}{p_\infty V_{g,0}} = \frac{1}{\gamma-1} \frac{p_{g,0}}{p_\infty}  \left( 1 - \left(\frac{V_g}{V_{g,0}}\right)^{1 - \gamma} \right)
    + \left(1 - \frac{V_g}{V_{g,0}} \right) - \frac{D_{\rm{eff}}}{p_\infty V_{g,0}},
    \label{eq:conservation_incomp}
\end{equation}
where $D_{\rm{eff}} > 0$ gathers all the mechanisms introducing an effective damping in the bubble oscillation. These effects include viscous dissipation, thermal difussion effects and the emission of outgoing pressure waves which act as an effective source of dissipation for open problems where the waves are not reflected back into the bubble.
This equation, applicable at every instant during the collapse and rebound process,
can be further simplified during the instants of minimum radius.
In the limit of large initial pressure differences ${p_\infty/p_{g,0}\gg 1}$,
the collapse is intense and the 
bubble volume is much smaller than the initial volume $V_g \ll V_{g,0}$ while the bubble pressure is 
much higher than the initial bubble pressure $p_{max} \gg p_{g,0}$. This allows us to simplify the equation above as
\begin{equation}
\label{eq:pmaxgeneral}
    \frac{p_{max}}{p_{g,0}} \approx \left[ 1 + (\gamma - 1) \frac{p_{\infty}}{p_{g,0}}  \left( 1 - \frac{E_{k,\rm{min}} + E_{s,\rm{min}}  + D_{\rm{eff},c}}{p_\infty V_{g,0}}\right) \right]^{\frac{\gamma}{\gamma - 1}}
\end{equation}
where $E_{k,\rm{min}}$ and $E_{s,\rm{min}}$ are the kinetic and surface energy at the instant of minimum volume and
$D_{\rm{eff},c}$ is the sum of the energy  dissipated by viscosity, the energy lost by heat diffusion and the energy emitted as an outgoing wave, all of them acting as penalization terms on the peak pressure reached during the collapse. \\

It is interesting to mention that the equation above is applicable for any arbitrary dimension and for any problem where the liquid pressure far from the bubble can be represented as constant,
as in the case of shock/bubble interaction
tests and underwater explosion simulations which are also widely reported in the literature. 
In the shock-wave/bubble interaction problem 
the bubble collapse 
large deformations are usually observed, which
reduces the peak pressures while the minimum volume attained
at the bubble collapse with respect to the idealized situation of a spherically symmetric Rayleigh collapse problem becomes larger. In addition,
some shockwave/bubble interaction tests are performed in 2D,
which introduces a correction factor on the determination of the minimum radius reached at the bubble collapse.

\subsubsection{Spherically symmetric limit}
It can be easily checked that for spherical bubbles in absence of dissipation eq. \ref{eq:pmaxgeneral} 
recovers the estimation provided by eq. \ref{eq:pinfsph}
by imposing that, for a spherical bubble in an incompressible liquid, the kinetic energy is null at the instant of minimum radius $E_{k,\rm{min}}=0$ and the surface energy negligible ($\frac{\sigma S_{I,0}}{p_\infty V_{b,0}} \ll 1$). Only when compressible effects are accounted for, the energy associated to 
the outgoing wave emitted during the collapse is not null,
acting as an effective energy lost included in $D_{\rm{eff}}$. Thus, for sufficiently intense collapses (e.g. $Re \to \infty$ and $We \to \infty$)  the preponderant mechanism introducing a correction on the peak pressures predicted by eq. \ref{eq:pinfsph} is liquid compressibility, reducing the peak pressures reached in the gas phase during the collapse.\\
\begin{figure}[h!]
    \centering
    \includegraphics[scale = 0.8]{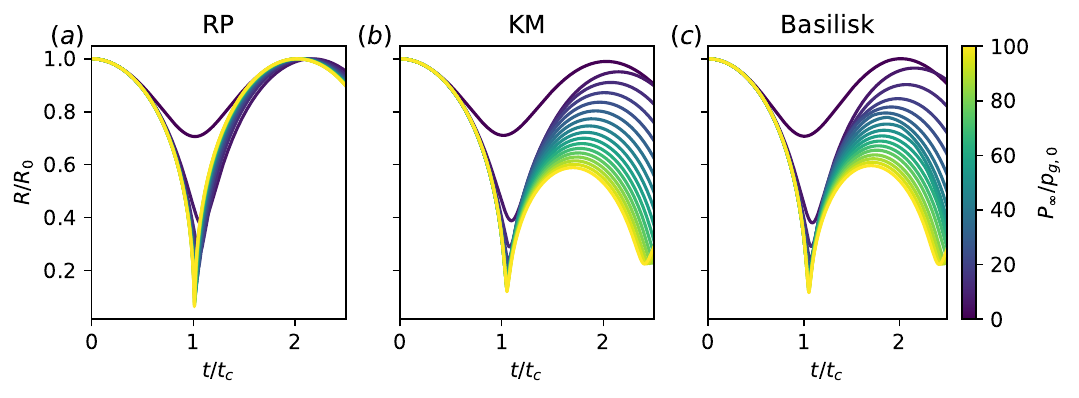} \\
    \includegraphics[scale = 1.25]{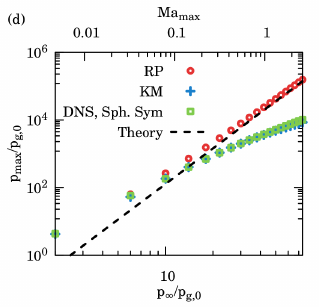}
    \caption{The bubble radius evolution as predicted from $(a)$ Rayleigh-Plesset model, $(b)$ Keller-Miksis model and $(c)$ Spherically-symmetric numerical simulation using Basilisk. Pressure ratio $p_\infty/p_{g,0}$ is varied to obtain different curves at $p_\infty/p_{g,0}$ for $p_{g,0} = 1 \textrm{atm}$, $c_l = 1500 m/s$ while neglecting effect of viscosity and surface tension $(\textrm{Re}_0 \to \infty, \textrm{We}_0 \to \infty)$. (d) The maximum gas pressure predicted from the Rayleigh-Plesset model, Keller-Miksis model, spherically symmetric numerical simulations and Equation \ref{eq:pinfsph}.}
    \label{fig:RPKM}
\end{figure}


Figure \ref{fig:RPKM}(a,b,c) shows the 
bubble radius evolution predicted from both Rayleigh--Plesset and Keller--Miksis models together with the DNS of the compressible Euler equations in both phases,
while figure \ref{fig:RPKM}(d) represents the maximum gas pressure during the collapse from these models along with equation \ref{eq:pinfsph}. As proposed by Saade et al. \cite{saade2023multigrid}, using the Mach number defined in equation \ref{eq:Mamax}, significant deviations with respect to the predictions for an incompressible liquid are observed for $\rm{Ma}_{\rm{max}} > 0.1$. The results of the DNS solver is in very good agreement with the predictions of the Keller--Miksis equation, which is consistent with other works 
contained in the literature for spherically
symmetric bubble collapses  \cite{fuster2011liquid,tiwari2013diffuse,schmidmayer2020assessment,denner2020modeling}.\\

\begin{figure}[h!]
\centering
\includegraphics[scale=0.9]{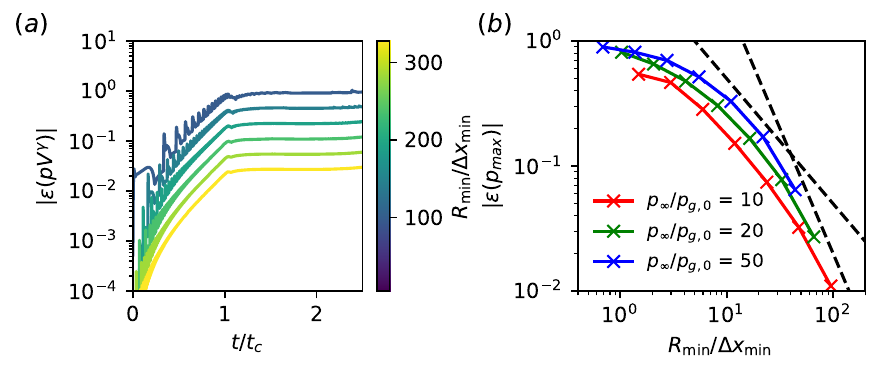}
\caption{\label{fig:CV_pVg} Spherically symmetric simulations. $(a)$ Temporal evolution of the relative error of $pV^\gamma$ for the collapse of a spherical bubble and
different grid sizes (color scale). $p_\infty/p_{g,0} = 20$.
$(b)$ Convergence analysis of the relative errors obtained in the prediction of the peak
pressures reached at the collapse for different intensities.
The errors scale well with the minimum radius reached for different collapse intensities.}
\end{figure}

The behavior of the numerical errors as a function of the grid resolution is shown in figure~\ref{fig:CV_pVg}.
Figure \ref{fig:CV_pVg}$a$ shows the relative change in $pV^\gamma$ during the collapse of spherical bubble in a spherically symmetric simulation for various grid resolution rescaled with minimum bubble radius $R_{min}/ \Delta x_{min}$
and for a constant value of the pressure ratio ($p_\infty/p_{g,0} = 20$).
As explained before, for an adiabatic collapse the errors
on the prediction of the maximum gas pressure depend on the capability of the numerical method to guarantee isentropic compression. 
The spurious non-isentropic behaviour is persistent near the minimum volume and these errors decrease very slowly with grid refinement. One requires as large as 50 grid points per minimum radius to get relative errors in the peak pressures down to $\mathcal{O}(10^{-1})$. The errors obtained for three different collapse intensities are shown to scale well with the minimum radius of the bubble
(figure~\ref{fig:CV_pVg}$(b)$), which for a uniform grid imposes the coarser resolution of the simulation. Remarkably, second order convergence
is shown to be achieved only when more
than 10 grid points per radius are used at the instant of minimum radius.
We can then conclude that reaching convergence is increasingly challenging as the intensity of the collapse increases as it requires the accurate 
resolution of the bubble at the instants of minimum radius
in order to obtain a correct estimation of the peak pressures generated at the bubble collapse.


\subsubsection{Non-spherical effects}

The validation of the predictions of the peak pressures reached in axisymmetric or full three dimensional simulations is not simple.
Even for bubbles initially spherical surrounded by a relative large amount of liquid, Rayleigh-Taylor instabilities during the collapse phase can result in a physically meaningful splitting and fragmentation of the bubble \cite{plesset1977bubble,brennen2002fission,Brennen}.
 The development of these instabilities are intrinsic to the problem and should not be artificially damped by the numerical diffusion of the numerical method, which at the same time must be accurate enough to avoid introduction of spurious non-spherical modes.\\

The influence of non-spherical deformations
is evident in cases 
where the bubble is initially not spherical or in problems
where an external asymmetry is responsible of the appearance of liquid jets \cite{naude1961mechanism,tomita1986mechanisms,johnsen2009,obreschkow2011universal}. A clear example is the investigation of the bubble collapse near a solid boundary \cite{saini2022dynamics} or a free surface \cite{blake1987cavitation,saade2021crown} which induce non-uniform perturbations around the bubble that can result in high speed liquid jets. As mentioned earlier, another related problem is the shock/bubble interaction problem where the shockwave is responsible of a significant deformation of the bubble during the collapse \cite{quirk1996dynamics}. In some cases, these jets can be very thin and extremely fast and hence require huge computational resources for proper resolution \cite{lechner2019fast,reuter2021supersonic}. The bubbles can also often take elliptic shapes too due to flow etc. \cite{starrett1983bubble,lechner2020jet,mnich2024single}.\\

\begin{figure}[hbt!]
    \centering
    \includegraphics[scale =0.75]{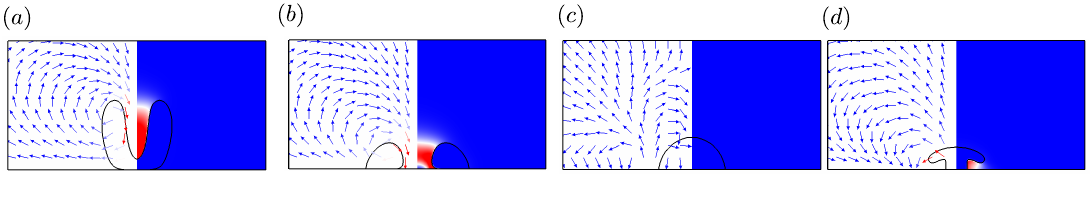}
    \caption{Snapshots of the bubble in contact with a wall at the instant of minimum volume for different initial values of initial contact angle $\alpha$. The interface is shown with black curve ,the kinetic energy in liquid phase is shown with linear color map in right half and the scaled velocity vectors in left half. The initial contact angle is $(a)$ $\alpha = 0$ $(b)$ $\alpha = \pi/3$ $(c)$ $\alpha = \pi/3$ $(d)$ $\alpha = 2/3\pi$}
    \label{fig:evol2}
\end{figure}

\begin{figure}[hbt!]
    \centering
    \includegraphics{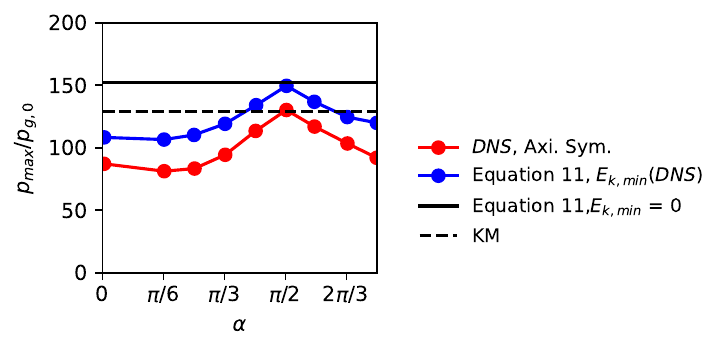}
    \caption{Axi-symmetric direct numerical simulations of the collapse of spherical cap bubbles in contact with a slip wall at ${p_\infty/p_{g,0} = 8}$,$\textrm{Re}_0 = \infty$, $\textrm{We}_0 = \infty$, $Ma_{max} = 0.117$ and varying contact angle. Peak pressure is shown as a function of contact angle from DNS (red curve),  solution of equation \ref{eq:pmaxgeneral} (solid black line) and the result of eq. \ref{eq:pmaxgeneral} introducing the kinetic energy obtained from DNS  at the instant of minimum radius (blue curve). The dotted black line indicates the predictions obtained from the KM model. 
    }
    \label{fig:kevsvol}
\end{figure}

\begin{figure}[h!]
    \centering
    \includegraphics{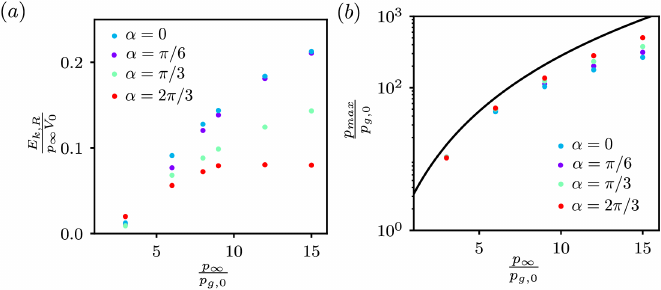} 
    \caption{DNS results for the collapse of spherical cap bubbles varying $p_\infty/p_{g,0}$  and the contact angle $\alpha$ at $\rm{Re}_0 = \infty$, $\rm{We}_0 = \infty$. $(a)$  Residual liquid kinetic energy at the instant of minimum volume. $(b)$ Peak pressures. The black curve corresponds to the upper bound of pressure predicted by equation \ref{eq:pinfsph}. }
    \label{fig:pmaxdeformed}
\end{figure}

From equation \ref{eq:pmaxgeneral} we can see that in the limiting case where $\rm{Re}_0 = \infty$, $\rm{We}_0 = \infty$ the peak pressures reached during the collapse are
penalized by liquid compressibility effects (already discussed
in the spherically symmetric case) and also by the presence of a residual kinetic energy at the instant of minimum radius.
The residual kinetic energy is directly attributed to the presence of vorticity in the flow, which, in the case of infinite Reynolds number, is represented
as a vortex sheet at the interface. 
Figure \ref{fig:evol2} shows snapshots at the instant of minimum volume for the simulations of bubble collapses for different
spherical cap bubbles in contact with a slip wall and 
in absence of viscous and surface tension effects.
The snapshots clearly show that the liquid is indeed not at rest
 at the instant of minimum radius, the magnitude of the kinetic energy being concentrated in the regions where liquid jets develop. 
The peak pressures
reached during the bubble collapse for $p_\infty/p_{g,0}=8$ 
and different values of the initial contact angle $\alpha$ 
(red curve in figure \ref{fig:kevsvol} ) reveal that the highest pressure is reached for the 90 degree angle case in which 
the bubble remains nearly spherical and $E_{K,R}\approx 0$.
In this case the peak pressure is close to the upper bound of the peak pressure predicted by eq. \ref{eq:conservation_incomp} (horizontal black line).
The peak pressures predicted from subtracting
the residual kinetic energy obtained numerically at the instant of minimum radius (blue line) predicts well the trend in pressure reduction due to the liquid kinetic energy. 
The consistent shift between the red and blue curves is a direct consequence of the damping due to liquid compressibility as confirmed in the case of spherically symmetric simulations, where the peak pressures are in agreement with Keller--Miksis predictions.
As expected, the influence of the contact angle on the peak pressures is increasingly important as the pressure ratio increases (figure \ref{fig:pmaxdeformed}), the peak pressures reached being significantly lower when the collapse is very intense (note that the plot is shown in log-scale) due to effects mentioned above. \\


\begin{figure}[h!]
    \centering
    \includegraphics[scale = 0.8]{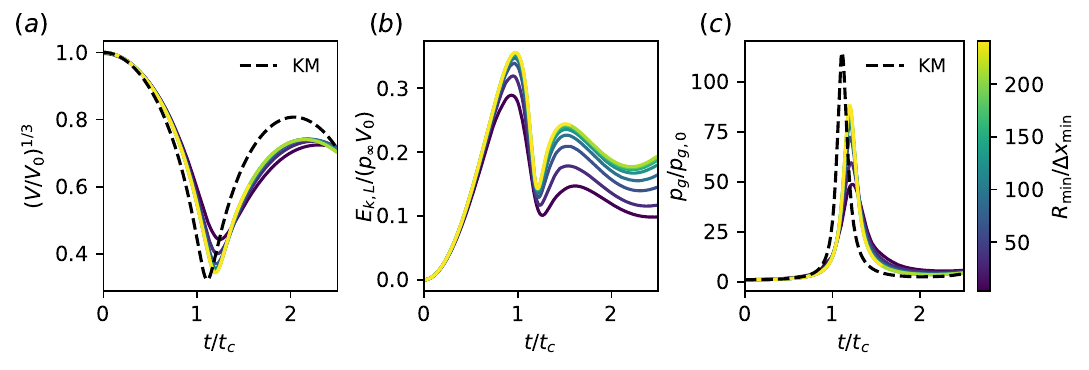} 
    \caption{Axi-symmetric simulations of Rayleigh collapse of a bubble close to a no-slip wall. Influence of the grid resolution on the temporal evolution of (a) bubble volume, (b) liquid kinetic energy $E_{k,l}$ and (c) the maximum gas pressure $p_{max}$. For this simulation, initial bubble size $R_0=0.5$mm,  $ p_\infty/p_{g,0} = 10$, and distance from the wall is  $d = 2R_0$. For reference, the solution for 
    a spherically symmetric bubble is also included.}
    \label{fig:convergence2}
\end{figure}


\begin{figure}[h!]
    \centering
    \includegraphics[scale = 1.]{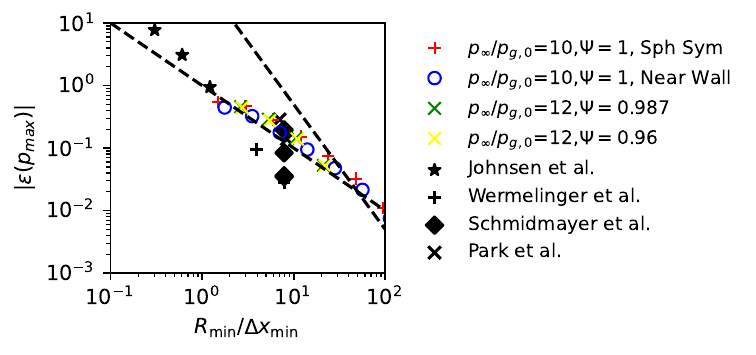}
    \caption{Relative error convergence on the maximal pressure predicted by the spherically symmetric simulations, axisymmetric simulations for the collapse of prolate bubbles with different values of the initial sphericity at $p_\infty/p_0 = 12$ and a 0.5mm air bubble in water collapsing at a distance $2R_{0}$ from a slip wall at $p_\infty/p_0 = 10$. An estimate of the relative error from references \cite{johnsen2009,wermelinger2018petascale,schmidmayer2020assessment,park2022numerical} is also shown with black symbols.}
    \label{fig:convergence3}
\end{figure}

The response of bubbles developing jets is similar for the collapse of a initially spherical bubble near a no-slip wall and some kinetic energy is present at the instant of minimum radius (figure \ref{fig:bubbleclosewall}). The grid convergence study for the evolution of kinetic energy integrated over the liquid control volume and the average bubble pressure for a representative case of $p_\infty/p_{g,0} = 10$ is shown in figures \ref{fig:convergence2}-\ref{fig:convergence3}, where we can see that the peak pressures and the amplitude of the rebound obtained from the convergence study differ significantly from the predictions of simplified spherically symmetric models. The overall convergence for these quantities is especially challenging at the instant of minimum kinetic energy and maximum pressure (minimum volume). We can see that the errors introduced in multidimensional simulations are consistent with those of spherically symmetric simulations despite the appearance of thin structures during the 
instants close to minimum volume which are poorly resolved in the case of axi-symmetric simulations as well as with the few studies available in the literature reporting the convergence of the peak pressures (e.g. minimum radius).\\ 

Finally, it is worth mentioning that secondary shock waves can be generated in non-spherical 
cases due to the impingement of the jet when piercing the bubble. The convergence of these secondary effects is controlled by the correct resolution of the peak jet velocities and has been shown to be sensitive to the viscous and surface tension effects that will be discussed next.



\subsection{Viscous and surface tension effects}
\begin{figure}[h!]
    \centering
    \includegraphics[scale = 0.9]{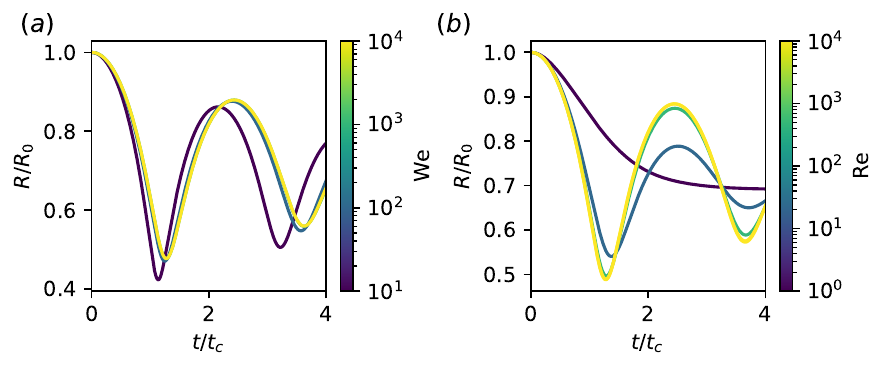}\\
    \caption{Evolution of bubble radius numerically obtained
    for a axi-symmetric collapse under $p_\infty/ p_{g,0} =5$ and varying $(a)$ Weber $\rm{We}_0$ at $\textrm{Re}_0 \to \infty$ $(b)$ Reynolds $\rm{Re}_0$ at $\textrm{We}_0 \to \infty$.}
    \label{fig:finite}
\end{figure}

The importance of viscous and surface tension effects
depends on the quantity of interest considered. 
Even for micron-sized bubbles,
if the bubble pressure at the instant of maximum radius 
is sufficiently close to the vapor pressure, large Reynolds and Weber numbers -- of the order of $\mathcal{O}(10^3)$ or larger -- 
have a negligible influence on the peak pressures and  the 
damping observed in 
spherically symmetric simulations as compared to the influence of compressibility effects (figure \ref{fig:finite}). \\


But viscous and surface tension can have an important
impact on the selection of additional length scales
associated to the bubble collapse.
One example is the aforementioned Rayleigh Taylor instabilities. 
Because the heavier fluid is accelerated towards lighter gas ($\dot{R} < 0, \ddot{R} > 0$), the collapse of a bubble is typically unstable and the interface develops non-spherical shapes \cite{Brennen}.
 At relatively large Reynolds numbers, the development and growth of the RT instabilities is coupled with the numerical errors and 
non-spherical effects become sensitive to the interface-capturing  and advection scheme \cite{schmidmayer2020assessment}. 
In this regime, the minimum grid size controls the 
sphericity of the interface.
When using more diffusive schemes or lower refinements, the non-spherical modes can be suppressed artificially which can be easily misunderstood as an improved result. \\

Viscous effects can be also relevant during the collapse of non-spherical bubbles. Particularly, the viscous effects may govern the dynamics of jet formation. The maximum velocity obtained during the collapse  sharply decays for small Reynolds numbers due to the suppression of the jet formation  \cite{popinet2002bubble,saini2022dynamics}.

\subsection{Rebound amplitude and damping}
\begin{figure}[h!]
    \centering
    \includegraphics[scale = 1.2]{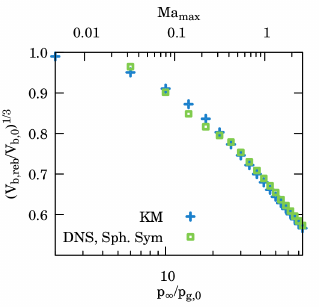}
    \caption{The rebound radius obtained from spherically symmetric simulations and from Keller-Miksis solution. $\rm{Re}_0=\infty$.
    $\rm{We}_0=\infty$, and varying $p_\infty/p_{g,0}$ and fixed sound speed in the liquid of $1500$m/s. }
    \label{fig:rreb}
\end{figure}

 The amplitude of the rebound is a quantity easily measurable experimentally. 
 This quantity provides us information about the
importance of the various mechanisms controlling the total effective damping which mainly act at extremely short lapses of time
during the bubble collapse, when the bubble is poorly resolved.
From equation \ref{eq:conservation_incomp} 
 and if we neglect
 the change of the surface energy with respect to the initial state, we obtain that
\begin{equation}
 \frac{V_{b,reb}}{V_{b,0}} \approx \left(1 - (\gamma - 1)  \frac{E_{k,reb} + D_{\rm{eff},reb}}{p_{g,0} V_{g,0}}   \right)^{\frac{1}{1 - \gamma}},
    \label{eq:vreb}
\end{equation}
which implies that the rebound amplitude is a direct measurement
of the residual energy contained in the liquid at the instant of maximum radius during the rebound and the total energy effectively dissipated during the collapse, this later quantity being controlled by the capability of the numerical method to capture the physics during the instant of minimum radius.
Figure \ref{fig:rreb} shows the amplitude of the rebound predicted by 
numerical simulations of the spherically symmetric case along with Keller-Miksis model predictions at $\textrm{Re} \to \infty$ and $\textrm{We} \to \infty$. Good agreement between DNS and Keller-Miksis predictions is observed confirming that liquid compressibility is the main damping mechanism influencing the amplitude of the rebound in this theoretical scenario.
However, this observation is in complete contradiction with recent experimental measurements reported by Preso et al \cite{preso2024vapour}, who have shown that the emission of a shock wave is not sufficient to explain the effective damping observed experimentally.
This represents a major challenge for numerical simulations
and models as these result reveal that liquid compressibility
effects are not the only relevant mechanism controlling the amplitude of the
rebound, the disagreement being especially relevant in the case of bubbles collapsing in water.
Note that many authors have reproduced the experimental evolution of the bubble radius using $p_{g,0}$ as a fitting
parameter.
However this procedure may hide fundamental
problems of the model and numerical method used as the value of $p_{g,0}$, which in reality is not a constant,
cannot be measured experimentally.
Only a careful experimental and numerical analysis about the sources of the damping can provide
insights about the correctness of the modelling equations and the numerical methods to predict the bubble response. Thus, the validation of a model and a numerical method requires to reproduce both the evolution of the bubble volume
and the energy released as an outgoing wave in order to ensure the accuracy of DNS predictions.
The disagreement between experimental observations and spherically symmetric models
observed by Preso et al \cite{preso2024vapour} can be attributed to bubble fragmentation, phase change, or dissociation. These mechanisms, typically neglected in numerical models used to reproduce experimental observations, may play a significant role on the effective damping observed experimentally.\\

\begin{figure}[h!]
    \centering
    \includegraphics[scale = 0.9]{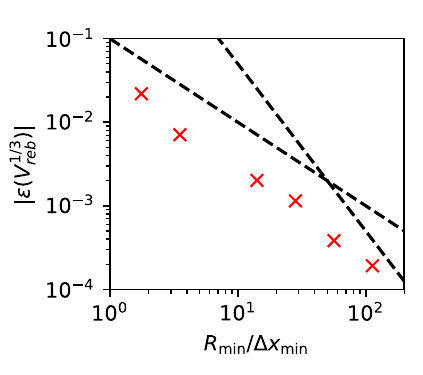}
    \caption{Relative convergence of the rebound volume for non-spherical collpase of 0.5 mm bubble under the influence of $p_\infty/p_{g,0} = 10$ at a distance $d=2 R_0$ from the no-slip wall.
    }
    \label{fig:rreb2}
\end{figure}

Finally, it is worth mentioning that similarly to the case of the peak pressures, 
the amplitude of the rebound also depends on the amount of kinetic energy in the liquid at the instant of the maximum radius.
This effect introduces additional effective damping
compared to the spherically symmetric case
which is translated into a lower amplitude of the rebound. This is clear in the cases of the collapse of a bubble close to a wall where the amplitude of the rebound is lower than in the case of spherically symmetric models (figure \ref{fig:convergence2}).
The convergence of this quantity is challenging and only first order convergence is achieved for the case considered 
even for a relatively high resolution of the interface (figure \ref{fig:rreb2}).\\


\subsection{Other relevant quantities}

The collapse time, the peak pressures reached in the gas phase and the rebound amplitude are examples of relevant quantities
that are shown to scale well with the minimum volume reached by the bubble
during the last instants of the bubble collapse.
However, in real problem applications, 
other quantities may be relevant.
For instance the adiabatic limit considered in many numerical models
holds for large bubbles when the time scales associated with the diffusion of heat across the interface are much larger than bubble collapse times. This regime corresponds to the limit of large Peclet numbers $\textrm{Pe} = \frac{R_0 U_c}{D^T_b} \gg 1$ where $D^T_b$ is the thermal diffusion coefficient in the bubble. In situations where the Peclet numbers is not sufficiently large, heat transfer effects can influence
the bubble response and need to be accounted for.
Thermal effects are also important 
to predict the ionization of gas inside the bubbles and the emission of light  \cite{gaitan1992sonoluminescence,brenner2002single}. The interface's temperature is usually very small compared to that of the bubble core, 
leading to the recombination of radicals 
produced at the bubble center that diffuse towards the cooled interface \cite{hauke2007dynamics}.
There are very few solvers capable of direct numerical simulations with thermal effects \cite{saade2023multigrid,bibal2024compressible,beig2015maintaining,beig2018temperatures}. 
An aditional important difficulty for the consideration of thermal effects is the absence of reliable Equation of States for both high pressure and temperature gases, which can reach plasma states. 
Apart from the use of a Nobel Abel stiffened gas equation of state, there are some models that are capable of using the thermodynamic tables for more accurate prediciton of thermal effects in bubble collapse \cite{bidi2023prediction}.
The correct treatment of thermal effects
is also the first step towards the correct implementation of phase change models \cite{bibal2024compressible,torres2024coupling}, given
that in many situations the heat diffusion in the thermal boundary layer around the bubble controls the total vaporization/condensation rate.\\

\begin{figure}[h!]
    \centering
    \includegraphics[scale = 0.8]{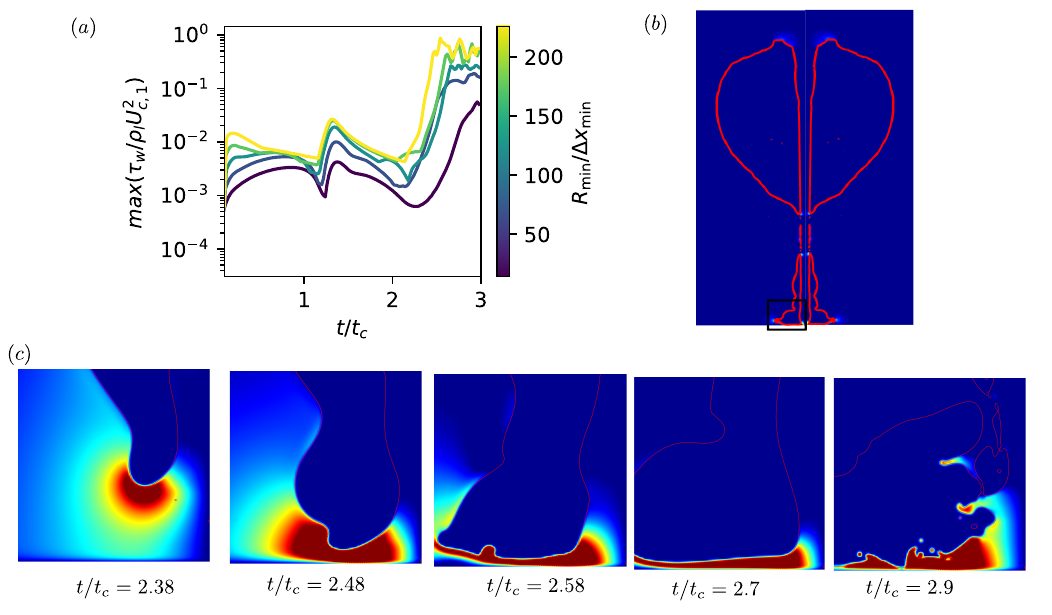}
    \caption{$(a)$ Evolution of maximum shear stress at the wall during the collapse of air bubble of initial radius $R_0 = 0.5$mm in water at distance $d=2 R_0$ from the wall for pressure ratio $p_\infty/p_{g,0} = 10$,  $\textrm{Re}_0 = 5000$ and $\textrm{We}_0 = 695$, $\rm{Ma}_{\rm{max}} = 0.085$. $(b)$ A zoom out of the bubble interface and region near the wall which is zoomed in panel $c$ is highlighted with black square. $(c)$ Velocity magnitude field near the no-slip wall close to the instant when the liquid jet reaches the wall and in the region highlighted in panel $b$.}
    \label{fig:tauw}
\end{figure}

\begin{figure}[h!]
    \centering    \includegraphics{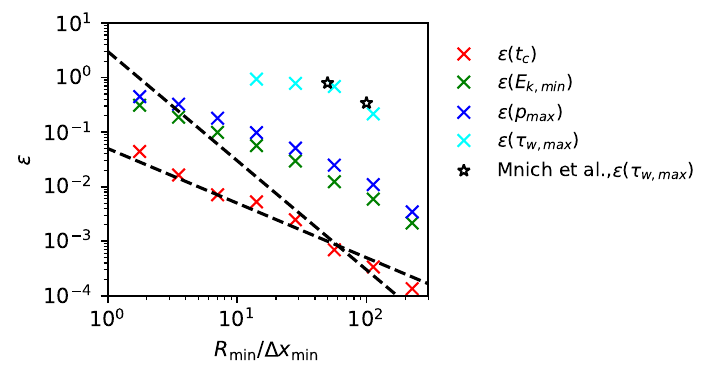}
    \caption{Grid convergence analysis for the collapse of 0.5mm air bubble in water at a distance $d=2R_0$ from wall.
    The relative errors are shown for the collapse time $t_c$, kinetic energy at minimum volume $E_{k,\textrm{min}}$, maximum gas pressure $p_{\textrm{max}}$ and maximum wall shear stress $\tau_{w,\textrm{max}}$. The numerical results from reference \cite{mnich2024single} are also added.}
    \label{fig:conv}
\end{figure}

Another relevant quantity for biomedical applications is the shear stresses induced on the solid during the collapse of a bubble nearby \cite{Zeng_2018,mnich2024single}. The interest of DNS is that it is not possible to access the values of shear stresses from experiments using conventional measurements techniques because the time scales associated are very fast. In figure \ref{fig:tauw}, we show the maximum shear stress during the collapse of 0.5 mm air bubble in water at distance $d = 2 R_0$ from the wall. The convergence of maximum shear stress is difficult to achieve due to the presence of poorly resolved jets in the coarse grid. As shown in figure \ref{fig:conv},
for this particular case the relative errors for the shear stresses are significantly larger than those obtained for any of the other quantities discussed previously, proving that in fully three-dimensional test
cases length-scales smaller than the minimum bubble radius control the accuracy of the numerical predictions. The use of Adaptive Mesh Refinement techniques certainly helps reducing the computational effort required to
resolve the appearance of thin structures during the collapse (figure \ref{fig:bubbleclosewall}), 
but most of currently available codes
do not allow for dynamic mesh adaptation which limits the capability 
of obtaining reliable predictions for strong bubble collapses.

\section{Conclusion}\label{sec13}

The development of robust and accurate numerical
methods for the Direct Numerical Simulation of bubble collapse processes
depend on the capability of quantifying the numerical errors introduced 
in the solution through numerical test cases.
Here we show that the minimum bubble volume, which can be used as a measurement of the collapse strength, controls the errors
introduced in a numerical solution. The dimensionless grid size obtained using the minimum length scale associated to the bubble collapse controls the errors of different relevant variables
including the collapse time, the peak pressures reached in the gas phase and the effective damping mechanisms acting during the 
last instants of the bubble collapse process.
A review of the existing grid convergence studies in
the context of bubble collapse problems show that currently available numerical methods require of at least 10 grid points per minimum radius to obtain relatively accurate results of the peak pressures reached during the collapse. This fact limits the conditions
that DNS can accurately reproduce with currently available computational 
resources. The lack of grid convergence studies 
for the shockwave/bubble interaction problem confirms that the large pressure ratios used experimentally in these problems prevent of an accurate numerical representation of the physical phenomena taking place inside the bubble at the instant of minimum radius.\\

An energy analysis for the problem of the Rayleigh collapse 
for a bubble of arbitrary shape shows that 
in addition to the acoustic
radiation of waves outwards of the bubble, the presence of kinetic energy
in the liquid effectively acts as a source of damping that reduces the peak
pressures reached during the collapse of the bubble and the amplitude 
of the rebound. These three dimensional effects cannot be captured by spherically symmetric models and simulations and require the resolution of 
the axi-symmetric or fully three dimensional problem. Some variables such as the viscous stresses exerted by
the bubble collapse on the wall or the development of Rayleigh--Taylor instabilities during the last instants of the bubble collapse may require additional efforts to resolve
the flow in characteristic length-scales that are smaller than the minimum equivalent bubble radius. The use of AMR techniques have been shown to be an effective method to reduce the errors on the prediction of these quantities.\\

\backmatter





\bmhead{Acknowledgments}
Part of this work was part of the PROBALCAV program supported by The French National Research Agency (ANR) and cofunded by DGA (French Minisitry of Defense Procurment Agency) under reference Projet ANR-21-ASM1-0005 PROBALCAV.
The authors would like to thank Antoine Llor and Stephane Zaleski for
insightful discussions.




\begin{appendices}

\section{Energy conservation equation for a bubble in a liquid}\label{secA1}

For a gas bubble with arbitrary initial shape inside a liquid bulk (figure \ref{fig:energy}) we can
consider a control volume for the bubble and 
another one for the liquid  moving with the local flow velocities. Let $V_g$ \& $V_l$ be the volume of bubble and liquid control volumes respectively, $S_g$ \& $S_l$ represent the surface area enclosing these control volumes and $\mathbf{n}_g$ \& $\mathbf{n}_l$ shows the unit normal to these surfaces pointing outward from the control volume. The total energy equation for either of the control volumes represented by the subscript $i \in (l,g)$ is
\cite{hauke2008introduction}
\begin{equation}
        \frac{d E_{e,i} }{dt}  + \frac{d E_{k,i}}{dt} = -\int_{S_i} p_i \mathbf{u}_i \boldsymbol{\cdot} \mathbf{n}_i dS 
        + \int_{S_i} \boldsymbol{\tau}_i' \mathbf{n} \cdot \mathbf{u}_i dS - \int_{S_i} \mathbf{q}_i \cdot \mathbf{n} dS,
    \label{eq:energygen}
\end{equation}
where the internal energy is defined as
$$E_{e,i} = \int_{V_i} \rho_i e_i dV.$$
Imposing that 
(i) the bubble expansion and collapse process is adiabatic and mass transfer effects are negligible,
(ii) the effect of body force terms (eg. gravity) is negligible and (iii) that the bubble pressure is uniform and well represented by an ideal gas law, 
the total energy conservation for the bubble can be expressed as
\begin{equation}
    \frac{p_{g,0} V_{g,0}^{\gamma}}{\gamma - 1} \frac{dV_g^{1 - \gamma}}{dt} + \frac{dE_{k,g}}{dt}= - p_b \frac{d V_g}{dt} - \frac{d D_g}{dt}.
    \label{eq:totenergygas}
\end{equation}

where $D_g$ captures all viscous and heat transfer processes across the interface, both of which can be eventually considered as energy lost mechanisms.
For the liquid phase, equation~(\ref{eq:energygen}) is expressed assuming that the velocity field is well represented by an incompressible velocity field plus a small
correction due to liquid compressibility effects
that effectively act as a damping mechanism included in $D_l$
\begin{equation}
        \frac{d E_{k,l}}{dt} = - \int_I \sigma \kappa \mathbf{u}_I \cdot \mathbf{n}_I dS + (p_b - p_\infty) \frac{dV_g}{dt} - \frac{dD_l}{dt}  - \frac{dE_{e,l}}{dt}.
    \label{eq:totenergyliq}
\end{equation}
Adding equations~(\ref{eq:totenergygas}) and~(\ref{eq:totenergyliq}) we readily obtain the total energy evolution equation for the system of a gas bubble in a liquid as 

\begin{equation}
    \frac{d E_{k}}{dt} + \frac{p_{g,0} V_{g,0}^{\gamma}}{\gamma - 1} \frac{dV_g^{1 - \gamma}}{dt} = - \int_I \sigma \kappa \mathbf{u}_I \cdot \mathbf{n}_I dS - p_\infty \frac{dV_g}{dt} - \frac{dD_{\rm{eff}}}{dt}
    \label{eq:energyboth}
\end{equation}
where $E_k = E_{k,l} + E_{k,g}$ and the effective damping of the system is defined including the internal energy of the liquid in order to account for the effective energy lost
due to acoustic radiation
$$\frac{dD_{eff}}{dt} = \frac{dD_g}{dt} + \frac{dD_l}{dt}  + \frac{dE_{e,l}}{dt}.$$
If $p_\infty$  is constant, we can integrate the equation~(\ref{eq:energyboth}) in time and obtain
\begin{equation} 
     E_{k} + E_s + \frac{p_{g,0} V_{g,0}^{\gamma}}{\gamma - 1} \left(V_g^{1 - \gamma} - V_{g,0}^{1 - \gamma}\right) + p_\infty \left(V_g - V_{g,0}\right) = - D_{\rm{eff}} + E_0,
    \label{eq:energyglbox}
\end{equation}
where $E_0$ is the integration constant which can be computed from the energy at the reference state
and 
$$E_s = \int_t \int_I \sigma \kappa \mathbf{u}_I \cdot \mathbf{n}_I dS dt$$
is the surface energy. Using 
 $$
 \int_I \kappa \mathbf{u}_I \cdot \mathbf{n}_I dS = - \int \nabla_s \cdot \boldsymbol{u} dS + \int \boldsymbol{u} \cdot \boldsymbol{p} dl
 $$
 with $\boldsymbol{p} = \boldsymbol{n} \times \boldsymbol{t}$ and $\boldsymbol{t}$ the tangent to the surface at the interface contour considered \cite{tryggvason2011direct}, 
it is easy to show that in the case of a bubble that is not in contact with the wall, the surface energy reduces to $dE_s = \sigma dS_I$, where $S_I$ is the total surface of the interface. Otherwise, the surface energy cannot be explicitly integrated and we need to account for the energy associated to the contact of both phases with the solid wall.


\end{appendices}

\bibliography{sn-bibliography}

\begin{mybox}{}{blue}
 \begin{wrapfigure}{l}{0.19\textwidth}
  \includegraphics[scale=0.18]{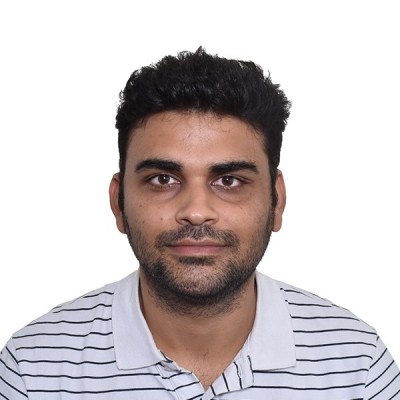}
  \end{wrapfigure}  
    \small \textbf{Mandeep Saini} Dr. Mandeep Saini obtained his PhD from the Sorbonne university. During his PhD he extensively studied the nucleation and collapse of bubbles using Basilisk solver. Before his PhD, he studied Bachelors and Masters in Mechanical Engineering from NIT Hamirpur and IIT Guwahati respectively. Currently, he is working at the University of Twente as a PostDoc researcher where he continues to develop numerical codes using Basilisk solver.\\\\

 \begin{wrapfigure}{l}{0.19\textwidth}
  \includegraphics[scale=0.08]{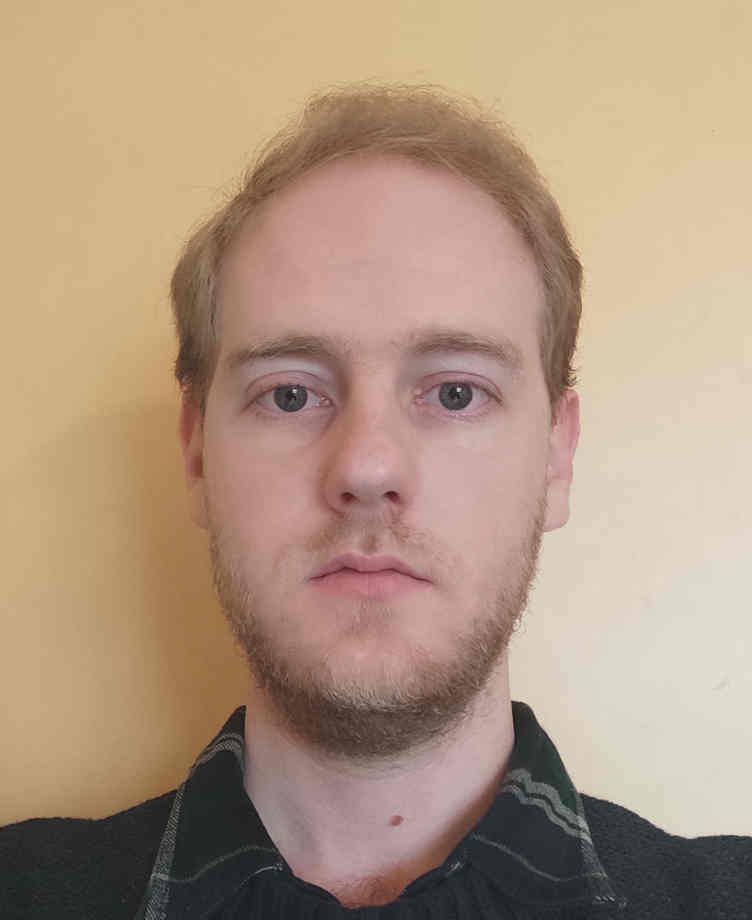}
  \end{wrapfigure}  
  
    \small \textbf{Lucas Prouvost} Dr. Lucas Prouvost obtained his PhD in Fluid Mechanics from Sorbonne University in 2022. He is an specialist in the development of numerical methods, with focus on Adaptive Mesh Refinement strategies and ALE formulations.\\\\

\begin{wrapfigure}{l}{0.19\textwidth}
  \includegraphics[scale=0.12]{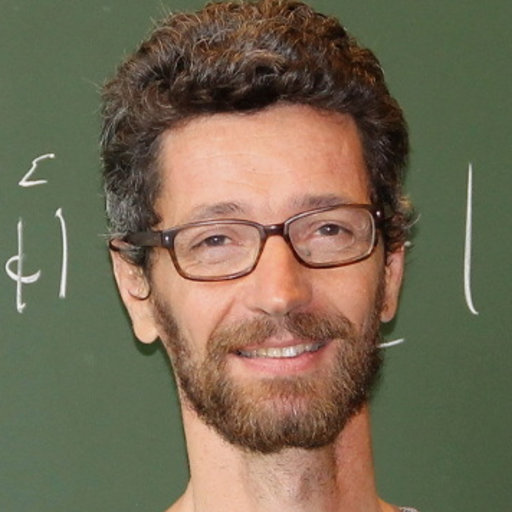}
  \end{wrapfigure}  
    \small \textbf{Stephane Popinet} Dr. Stephane Popinet
    is Directeur de recherche CNRS at the Institut Jean le Rond d'Alembert (Sorbonne Université).
Stéphane Popinet's work focuses on the development of numerical methods in fluid mechanics with applications in tsunamis, multiphase flows in presence of surface tension and granular media. He is the author of the free-softwares Gerris and Basilisk.\\\\

 \begin{wrapfigure}{l}{0.19\textwidth}
   \includegraphics[scale=0.025]{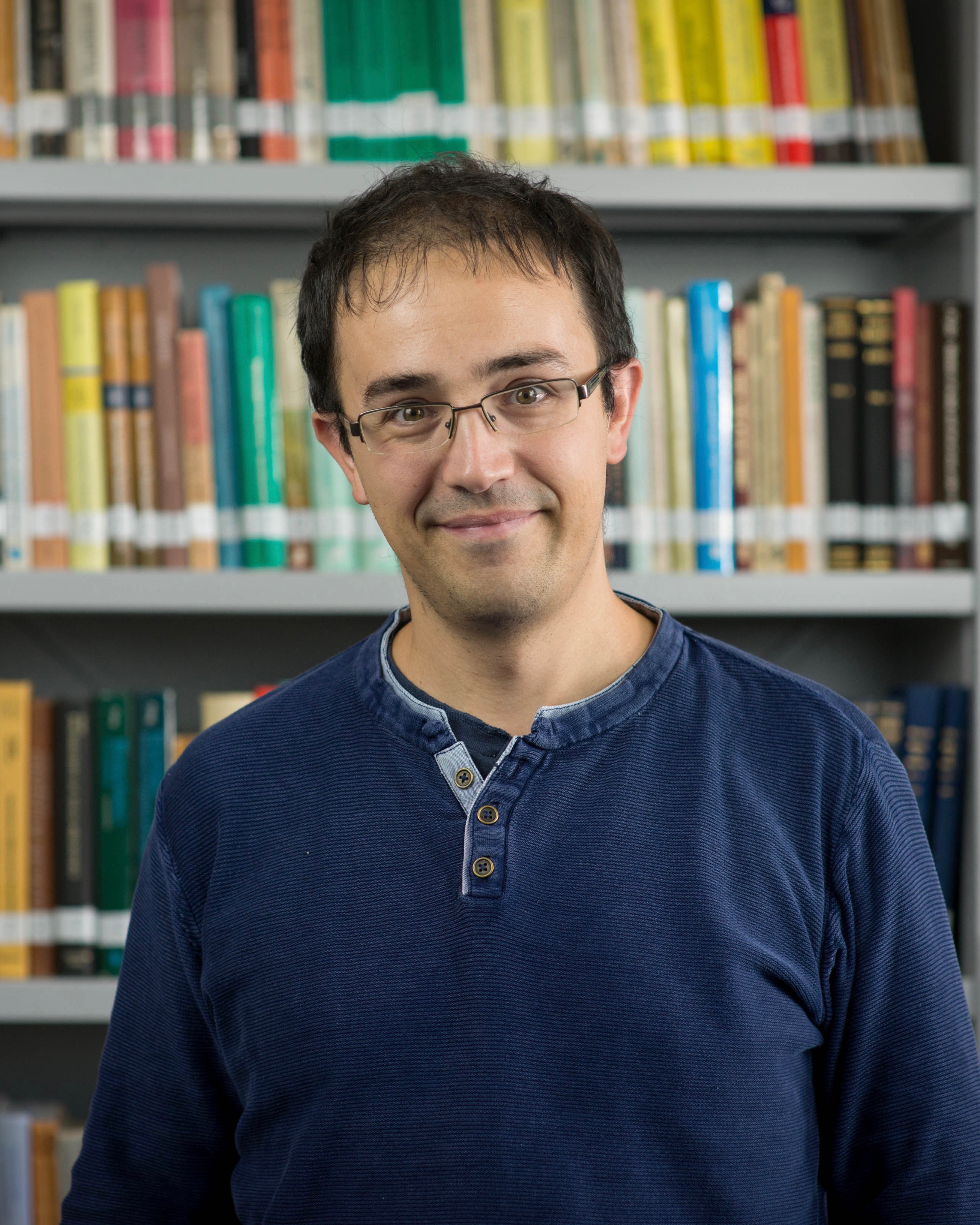}
   \end{wrapfigure}  
\small \textbf{Daniel Fuster} Dr. Daniel Fuster received a
Ph.D. degree in fluid mechanics in 2007 from University of Zaragoza. He held
post-doctoral research positions at UPMC (Paris)
and Caltech (USA) until 2010, when he became a
permanent CNRS researcher in the Institut Jean Le
Rond D’Alembert at Sorbonne University (Paris)
where he is currently Directeur de Recherche at CNRS.
His main area of interest is the development of
models and tools for the numerical simulation of
multiphase flows, focusing in particular on the individual and collective response of bubbles and bubble
clusters.
\end{mybox}

\end{document}